\documentstyle[amssymb,aps,prl,psfig,twocolumn]{revtex}

\begin{document}
\draft
\title{Cosmic implications on thermodynamics\\
and the explanation of the so called horizon problem}
\author{Osvaldo M. Moreschi\thanks{%
Member of CONICET.}}
\address{FaMAF\\
Universidad Nacional de C\'{o}rdoba\\
Ciudad Universitaria\\
(5000) C\'{o}rdoba, Argentina\\
{\small Email: moreschi@fis.uncor.edu}}
\maketitle

\begin{abstract}
We show that there are implications on thermodynamics that come from the
existence of the initial cosmic singularity. At present time this is more a
conceptual change than an observable one. However at very early cosmic times
there is a big difference between the actual behavior of thermodynamic
quantities and the behavior assumed in the standard cosmological model. We
present the discussion of two systems: an ideal monatomic gas at present,
and a photon gas at the early Universe. We show the striking result that the
entropy density goes to zero as the cosmic time goes to zero. This in turn,
provides an explanation for the so called horizon problem.
\end{abstract}

\pacs{PACS number(s): 98.80.C, 05.20, 65.50 }


When one studies a thermodynamic system one normally have the thermodynamic
laws, resting on our knowledge of the phenomenological equations of state;
and one also seeks for an explanation of all these laws from statistical
mechanics. If for example we study a simple system like a {\bf classical
ideal monatomic gas} composed of $N$ molecules in a volume $V$ at
temperature $T$; we have the phenomenological thermodynamic description in
terms of the equations of state for the pressure $P$ and the internal energy 
$U$; namely $P_{{\em Ph}}=(Nk_{B}T)/V$ and $U_{{\em Ph}}=\frac{3}{2}Nk_{B}T$%
\cite{bCallen85}; where $k_{B}$ is Boltzmann's constant and we are using the
subscript $_{{\em Ph}}$ to emphasize that these expressions have a
phenomenological origin. One also has the explanation of these equations
from a microscopical description from statistical mechanics. In the process
of performing a calculation in the framework of statistical mechanics, one
needs to settle the appropriate distribution of probabilities for the
microscopic states. This distribution is normally calculated from the
maximum entropy principle\cite{bCallen85}\cite{bLandaustat80}; which
basically states that the probabilities $p_{i}$ for the system to be in the
microscopic state $i$ is to be calculated by maximizing the entropy $%
S(U,V,N) $ subject to the information that we have on the system. For
example in the above case of the classical ideal gas, one would say that
while the volume $V $ and the number of molecules $N$ are fixed; we only
know for the microscopic energy that its average value must coincide with $U$%
; that is, $\sum p_{i}E_{i}=U$; since the fact that our system is at
temperature $T$, precludes us from a more precise information on the energy
of the system. Therefore to calculate the distribution of probabilities $%
\left\{ p_{i}\right\} $ we must maximize the entropy 
\begin{equation}
S=-k_{B}\sum p_{i}\ln p_{i}  \label{eqentropy1}
\end{equation}
subject to the condition 
\[
\sum p_{i}E_{i}=U. 
\]

The distribution so obtained is given by: 
\begin{equation}
p_{i}=\frac{e^{-\beta E_{i}}}{\sum_{j}e^{-\beta E_{j}}},  \label{eprobabi}
\end{equation}
where $\beta =1/(k_{B}T)$; which is recognized as the Gibbs or canonical
distribution.

We have said before that the entropy must be maximized subject to the
information that we have on the system; and we do want to take into account
that we live in a Universe with an initial cosmic singularity. The way in
which this observation affects our statistical mechanic calculation is by
imposing some constrains on the available phase space of our system. More
precisely, the existence of an initial cosmic singularity implies that there
are particle horizons\cite{bRindler56}; which means that our past and the
amount of matter that we can observe are finite. The implication of this is
that we can not assume that our thermodynamic system is in contact with an
infinite reservoir at temperature $T$. In particular this means that not all
of the theoretically predicted energy levels $E_{j}$ are available; since,
if we label with $E_{\max }$ the total available energy in our causal past,
then clearly we can only consider $E_{j}\leq E_{\max }$.

Following the standard calculation for a classical ideal gas\cite{bCallen85}
we find that the partition function $Z=\sum_{j}e^{-\beta E_{j}}$, appearing
in the denominator of (\ref{eprobabi}) above, is given by

\begin{equation}
Z=\sum_{j,j\prime ,j\prime \prime ,...}e^{-\beta \left( \varepsilon
_{1j}+\varepsilon _{2j\prime }+\varepsilon _{3j\prime \prime }+...\right) };
\label{eqpartition1}
\end{equation}
where $\varepsilon _{ij}$ is the energy of the particle $i$ in the state $j$%
. In the usual presentation of a classical ideal gas, it is assumed
non-interaction and statistical independence for each atom of the gas; from
which it is deduced that the sums over each of the indices appear as
factors; so that the partition function would be given by $Z=z^{N}/N!$,
where $z$ is given, in the continuous representation, by

\begin{equation}
z=\frac{V}{h^{3}}\int e^{-\beta \frac{p^{2}}{2m}}dp_{x}dp_{y}dp_{z};
\label{eqz1}
\end{equation}
where $h$ is Planck constant. The correcting factor $1/N!$, accounts for the
indistinguishability of the particles\cite{bCallen85}\cite{bLandaustat80}.

Due to the restriction on the phase space coming from our cosmological
knowledge we are not free to integrate in (\ref{eqz1}) up to arbitrary large
values of $p^{2}$; since $\frac{p^{2}}{2m}$ can not be larger that the total
available energy $E_{\max }$ in our causal past; in other words we must have 
$\frac{p^{2}}{2m}\leq E_{\max }$. Therefore, one has $z=\frac{\pi V}{h^{3}}%
\sqrt{2^{5}\left( m/\beta \right) ^{3}}\int_{0}^{\beta E_{\max
}}e^{-x}x^{1/2}dx$. In the limit for $E_{\max }$ going to infinity one
obtains the usual result $z_{\infty }=(V/h^{3})(2\pi mk_{B}T)^{3/2}$;
however due to the cosmological restriction we can show that

\begin{eqnarray*}
\ln Z(\beta ,V,N) &=&N\ln \left( \frac{\pi V}{h^{3}}\sqrt{2^{5}\left(
m\right) ^{3}}\right) -\frac{3}{2}N\ln \beta \\
&&+N\ln \int_{0}^{\beta E_{\max }}e^{-x}x^{1/2}dx-\ln N!\quad .
\end{eqnarray*}

Then the entropy is given by

\[
S=k_{B}\beta \sum \frac{e^{-\beta E_{i}}}{Z}E_{i}+k_{B}\ln Z=k_{B}\beta
U+k_{B}\ln Z. 
\]
%
And the corresponding energy is given by

\begin{equation}
U=Nk_{B}T\left( \frac{3}{2}-\frac{x_{c}^{3/2}e^{-x_{c}}}{%
\int_{0}^{x_{c}}e^{-x}x^{1/2}dx}\right)  \label{eqUgas1}
\end{equation}
with $x_{c}\equiv \beta E_{\max }$. To give an idea of the order of
magnitude of the correcting term, let us consider $T=300{^{\circ }}{\rm K}$,
and a cosmic mass density $\rho $ to be two percent of the critical mass
density; that is, $\rho =0.02\rho _{c}=(0.02)3H_{0}^{2}/8\pi G$, with the
Hubble parameter $H_{0}$ with value $H_{0}=75(km/s)/Mpc$; and where $G$ is
the gravitational constant. Assuming a sphere of radius $R=c/H_{0}$, where $%
c $ is the velocity of light, one can estimate $E_{\max }$ by $E_{\max
}=4\pi R^{3}\rho c^{2}/3$; which implies a value of $x_{c}$ on the order of $%
x_{c}\simeq 10^{88}$. Such huge value for $x_{c}$ gives a negligible
correction in front of $3/2$ in equation (\ref{eqUgas1}). Even if one
considers temperatures as high as those found in the center of the sun, one
will have $x_{c}\simeq 10^{83}$; and the whole correcting term to the
equation of state (\ref{eqUgas1}), would be of order $10^{-125}$ in front of
the $3/2$.

The pressure can be calculated from

\[
P=\frac{1}{\beta }\frac{\partial \ln Z}{\partial V}=\frac{N}{\beta V}; 
\]
therefore in this case the equation of state for the pressure does not show
any correction; that is $P=P_{{\em Ph}}$.

It is observed that in the previous expressions the correcting term is more
important if $\beta E_{\max }$ becomes smaller. This situation arises when
one approaches the initial cosmic singularity; since in this regime $\beta $
goes to zero and also $E_{\max }$ shows the same behavior; due to the fact
that the particle horizon reduces its size as one considers earlier cosmic
times.

At very early cosmic times, during the radiation dominated era, the
different kind of particles contribute to the energy momentum tensor as
different components of a relativistic gas. In order to be concrete we will
study next the system corresponding to a {\bf photon gas} in a small volume $%
V$ at temperature $T$, at very early cosmic times.

The corresponding phenomenological equations of state for this system 
are: $U_{{\em Ph}%
}=a\;V\;T^{4}$ and $P_{{\em Ph}}=\frac{1}{3}U_{{\em Ph}}/V\,$; where $%
a=4\sigma _{0}/c$, and $\sigma _{0}$ is the Stefan-Boltzmann constant. From
the knowledge of the equations of state, one can calculate the
phenomenological entropy; which is given\cite{bCallen85} by $S_{{\em Ph}}=%
\frac{4}{3}a^{1/4}\;U_{{\em Ph}}^{3/4}\;V^{1/4}$.

In a Friedman Universe the line element is given in terms of the metric of a
Robertson-Walker spacetime, namely\cite{bWald84}

\begin{equation}
ds^{2}=dt^{2}-\left( \frac{A(t)}{c}\right) ^{2}dl_{k}^{2};
\label{linelement}
\end{equation}
where $A(t)$ is the expansion parameter with units of length and $dl_{k}^{2}$
is the line element of a homogeneous and isotropic space with constant
curvature $k=1,0$ or $-1$. In a radiation dominated Universe, $\rho \,A^{4}$
is a constant; and for very early cosmic times, $A(t)$ behaves
asymptotically as $\sqrt{t}$; where the initial singularity is attainned in
the limit $t\rightarrow 0$. Then, one can deduce that the phenomenological
entropy density, $s_{{\em Ph}}\equiv S_{{\em Ph}}(V_{0})/V_{0}$, diverges as 
\fbox{$s_{{\em Ph}}\varpropto t^{-\frac{3}{2}}$} when $t\rightarrow 0$.

Let us now proceed with the correct causal statistical mechanic calculation
of the entropy, taking into account our knowledge of the initial
singularity. As we have seen in the previous example, we expect to find a
different expression for the correct entropy; which implies that the
corresponding equations of state will be different to those that appear in
the standard cosmological model dominated by radiation. This means that to
be absolutly consistent one should solve the Einstein equations taking this
effect into account. We plan to do this in a future paper; instead here we
would like to concentrate in the general behavior of the entropy in a
Universe with an initial cosmic singularity, by given the calculation in a
fixed geometry. We choose this geometry to be the Friedman line element
corresponding to a Universe dominated by radiation. We also have the freedom
to relate one of our thermodynamic variables to the geometry; we do this by
demanding the energy density to behave as $1/A^{4}$; as it does in the
standard cosmological model of a Universe dominated by radiation.

In order to be explicit, let us consider an open Universe; that is, the $%
k=-1 $ case, whose expansion parameter is deterimined from the Friedmann
equations\cite{bWald84}

\begin{equation}
A(t)=\left( \frac{\sqrt{Q_{r}}}{c}\right) \sqrt{2\left( \frac{tc^{2}}{\sqrt{%
Q_{r}}}\right) +\left( \frac{tc^{2}}{\sqrt{Q_{r}}}\right) ^{2}},
\label{Adet}
\end{equation}
and where the density $\rho $ is linked to the geometry in such a way that
the cuantity $Q_{r}$, defined by $Q_{r}\equiv (8\pi G/3)\rho A^{4}$, is a
constant.

Let us consider a reference volume $V_{0}$, which is small enough so that
any other dynamical time is much larger than the typical thermodynamic
relaxation time\cite{bLandaustat80} associated to our subsystem determined
by $V_{0}$. An observer that is comoving with the matter flow ascribes the
entropy $S(t)=S(U(t),V_{0})$ to this small subsystem; where the world line
of the observer can be parametrized by the cosmic time $t$. Let us recall
that the chemical potential for a photon gas is zero; therefore the entropy
does not depend on the number of particles.

%
%

Following the standard procedures appearing in text books\cite{bHuang} of
statistical mechanics one can prove that the entropy and internal energy for
electromagnetic radiation contained in a volume $V_{0}$ are given, 
in the continuum representation, by

\begin{eqnarray}
S &=&\frac{V_{0}k_{B}}{\pi ^{2}c^{3}\beta ^{3}\hbar ^{3}}\left( \left. -\ln
\left( 1-e^{-w}\right) \frac{w^{3}}{3}\right| _{0}^{x_{\max }}\right.
\label{entropycont2} \\
&&\left. +\frac{4}{3}\int\limits_{0}^{x_{\max }}\left[ \frac{w^{3}}{\left(
e^{w}-1\right) }\right] \;dw\right) \,  \nonumber
\end{eqnarray}
and

\begin{equation}
U=\frac{V_{0}}{\pi ^{2}c^{3}\beta ^{4}\hbar ^{3}}\int\limits_{0}^{x_{\max
}}\left[ \frac{w^{3}}{\left( e^{w}-1\right) }\right] \;dw\,;
\label{eqUcosmic1}
\end{equation}
where $x_{\max }(t)=\beta (t)\ E_{\max }(t)$ and the dimensionless variable
of integration $w$ is defined by $w=\beta \hbar \omega $.

If there where no particle horizons the first term would not contribute to
the value of the entropy; since in this case $x_{\max }=\infty $. Then, in
this situation one would reproduce the expression for $S_{{\em Ph}}$
mentioned above.

It is important to remark that the Stefan-Boltzmann law; which gives the
energy radiated per second per unit area $I$ for black body radiation,
namely, $I=(cU_{{\em Ph}})/(4V)=\sigma _{0}\;T^{4}$, is now corrected by the
expression (\ref{eqUcosmic1}). One way to understand this result is that the
Stefan-Boltzmann constant $\sigma _{0}$, mentioned above, is actually a
parameter that changes with cosmic time; which is given by $\sigma (t)=\frac{%
k_{B}^{4}}{4\pi ^{2}c^{2}\hbar ^{3}}\int\limits_{0}^{x_{\max }(t)}\left[ 
\frac{w^{3}}{\left( e^{w}-1\right) }\right] \;dw$; and where one usually
approximates for the present time $t_{0}$ by the value $\sigma
(t_{0})\backsimeq \sigma _{0}\equiv \frac{k_{B}^{4}}{4\pi ^{2}c^{2}\hbar ^{3}%
}\int\limits_{0}^{\infty }\left[ \frac{w^{3}}{\left( e^{w}-1\right) }\right]
\;dw=\frac{\pi ^{2}k_{B}^{4}}{60c^{2}\hbar ^{3}}$.

\begin{figure}[h]
\psfig{file=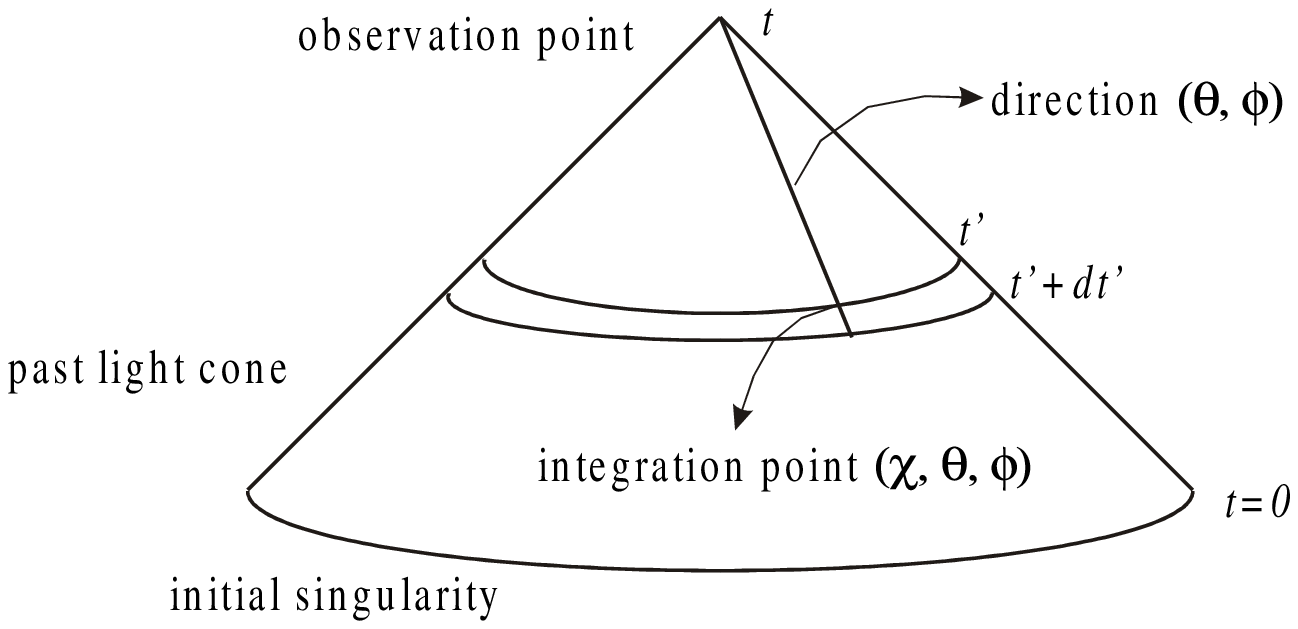,width=8.6cm}
\caption{The observation point is the tip of the past light cone with the
generic time value $t$. A direction of observation with angular coordinates $%
(\theta ,\phi )$ reaches the sphere in the past at time $t^{\prime }<t$,
with radial coordinate $\chi $. The past light cone extends up to the
initial cosmic singularity with the time value $t=0$.}
\label{cono}
\end{figure}

When constructing a cosmological model, one normaly represent matter by some
kind of fluid. In any Universe whose energy momentum tensor is that
corresponding to a fluid, given an event $p$, one can calculate the total
available energy in the causal past of $p$ by integrating the contribution
of the fluid on the boundary of the causal past of $p$. In our case we can
calculate $E_{\max }(t)$ explicitly by considering the contributions on the
past light cone at earlier times $t^{\prime }$ coming from the energy
density $\rho (t^{\prime })c^{2}$, with the corresponding red shift factor $%
A(t^{\prime })/A(t)$ and multiplied by the volume element $4\pi \
r(t,t^{\prime })^{2}A(t^{\prime })\ dl=4\pi \ r(t,t^{\prime })^{2}c\
dt^{\prime }$; where $r(t,t^{\prime })$ is the radius of a sphere, with
center at the point of observation at cosmic time $t$, and with surface on
the past light cone at time $t^{\prime }$. Considering the realistic case $%
k=-1$, corresponding to a Universe with energy density below the critical
value, the line element of the three dimensional homogeneous and isotropic
space can be expressed by $dl^{2}=d\chi ^{2}+\sinh (\chi )^{2}(d\theta
^{2}+\sin (\theta )^{2}d\phi ^{2})$. Using these coordinates, the radius $%
r(t,t^{\prime })$ of the sphere is given by $r(t,t^{\prime })=A(t^{\prime
})\sinh (\chi (t,t^{\prime }))$; where the center of the coordinate system
is taken at the point of observation, so that $\chi (t,t)=0$. In the
spacetime diagram in Fig. \ref{cono} one can see a picture of the coordinate
system.

One can calculate $\chi (t,t^{\prime })$ from the
expression $\chi (t,t^{\prime })=c\int_{t^{\prime }}^{t}dt^{\prime \prime
}/A(t^{\prime \prime })$; from which it is obtained $E_{\max }(t)$, namely

\begin{eqnarray*}
E_{\max }(t) &=&\int_{0}^{t}\rho (t^{\prime })c^{2}\frac{A(t^{\prime })}{A(t)%
}4\pi r(t,t^{\prime })^{2}c\ dt^{\prime } \\
&=&\frac{3Q_{r}c^{2}}{4G}\frac{1}{A(\tau )}\left[ \left( \tau +1\right) 
\sqrt{\tau ^{2}+2\tau }\right. \\
&&\left. -\ln (\tau +1+\sqrt{\tau ^{2}+2\tau })\right] ;
\end{eqnarray*}
where we are using the convenient dimensionless time variable $\tau =(tc^{2}/%
\sqrt{Q_{r}})$. In order to study the behavior of $E_{\max }(t)$ near the
initial singularity it is useful to express it in terms of $A(\tau )$; which
is also a good time variable, since it is a monotonically increasing
function of time. In this way one obtains

\begin{eqnarray}
E_{\max } &=&\frac{3\sqrt{Q_{r}}c^{3}}{4G}\left[ \left( \sqrt{\left( \frac{c%
}{\sqrt{Q_{r}}}A\right) ^{2}+1}\right) \right.  \label{maxenergy} \\
&&\left. -\frac{\sqrt{Q_{r}}}{cA}\ln \left( \sqrt{\left( \frac{c}{\sqrt{Q_{r}%
}}A\right) ^{2}+1}+\frac{c}{\sqrt{Q_{r}}}A\right) \right] .  \nonumber
\end{eqnarray}

It is clear that as $t$ decreases, so does the region covered by the
particle horizon; and therefore $E_{\max }(t)$ also decreases. More
specifically one can show that the asymptotic expression of $E_{\max }$ for
small values of $A$ starts with: $E_{\max }(A)=(3\sqrt{Q_{r}}%
c^{3}/4G)[(2/3)(cA/\sqrt{Q_{r}})^{2}+O((cA/\sqrt{Q_{r}})^{4})]$. In
particular, one can see that in the limit $A\rightarrow 0$, one has

\begin{equation}
\lim_{A\rightarrow 0}E_{\max }=0\,.  \label{energylimit}
\end{equation}


It should be emphasized that although the energy per unit volume diverges as
one approaches the initial cosmic singularity, the maximum possible
available energy at any given point is finite, and it goes to zero as $%
t\rightarrow 0$.

In order to study the behavior of the entropy per unit volume in the
vicinity of the initial singularity, we can make an asymptotic expansion of
the entropy for very small values of $t$. Defining the entropy density by $%
s(t)\equiv S(t)/V_{0}$, one obtains

\begin{eqnarray}
s(t) &=&\frac{k_{B}}{\pi ^{2}c^{3}\hbar ^{3}}\left( -\frac{E_{\max }(t)^{3}}{%
3}\ln \left( \beta (t)E_{\max }(t)\right) \right. \,  \label{entropyasymp2}
\\
&&\left. +\frac{4}{9}E_{\max }(t)^{3}\right) +O(\beta (t)E_{\max }(t)^{4}). 
\nonumber
\end{eqnarray}

While in the standard model $\beta (t)$ is given by $\frac{\beta (t_{0})}{%
A(t_{0})}A(t)$; from our setting one can deduce that actually, $\beta (t)$
must satisfy

\begin{equation}
\beta (t)=\frac{\beta (t_{0})}{A(t_{0})}A(t)\left[ \frac{\int\limits_{0}^{%
\beta (t)\ E_{\max }(t)}\left[ \frac{w^{3}}{\left( e^{w}-1\right) }\right]
\;dw}{\int\limits_{0}^{\beta (t_{0})\ E_{\max }(t_{0})}\left[ \frac{w^{3}}{%
\left( e^{w}-1\right) }\right] \;dw}\right] ^{\frac{1}{4}};  \label{beta}
\end{equation}
which reduces to the usual behavior when $\beta (t)\ E_{\max }(t)$ and $%
\beta (t_{0})\ E_{\max }(t_{0})$ are big. Instead, when $\beta (t)\ E_{\max
}(t)$ is small, this expression shows a very different behavior for $\beta
(t)$. From the fact that $E_{\max }(t)$, for very small $t$, behaves as $%
A(t)^{2}$, one can deduce from equation (\ref{beta}) that $\beta (t)$
behaves as $A(t)^{10}$, in the vicinity of the initial singularity.

Using both asymptotic behaviors; namely: $E_{\max }(t)\varpropto A(t)^{2}$
and $\beta (t)\varpropto A(t)^{10}$, as $t$ goes to zero, it is easily seen
from equation (\ref{entropyasymp2}) that the entropy density goes to zero
when one approaches the initial cosmic singularity; that is

\begin{equation}
\fbox{$\ \lim\limits_{t\rightarrow 0}s(t)=0\ $.}  \label{entropylimit}
\end{equation}


The consequences of our result regarding the possible initial data for
physical fields at the very early Universe, can be better understood if we
recall from equation (\ref{eqentropy1}) that the entropy is given by $%
S=-k_{B}\sum\limits_{i}p_{i}\ln p_{i}$; since this expression implies that
the only way in which the entropy of a system can go to zero is if all the
probabilities $p_{i}$ go to zero except one, let us say $p_{0}$, which must
have the unit value. This means that if we where to have two subsystems with
identical statistical mechanic description and with zero entropy, then the
physical state of both systems would be indistinguishable from the
statistical mechanic point of view. This result explains way when we observe
the cosmic background radiation in different directions on the sky that
where causally disconnected at the time of decoupling, have such a similar
description

The attempts to understand the low entropy behavior at the early Universe
has been the subject of study from different viewpoints. For example the
standard presentations of the postulated inflationary Universes aims to
explain the so called horizon problem, which is associated to the low
entropy behavior\cite{Guth81}.

According to our result we see that there is no problem with the observed
initial isotropic and homogeneous behavior of the Universe, regarding the
existence of the initial cosmic singularity; since we have seen that such
behavior is actually a consequence of the presence of the cosmic singularity.

\acknowledgments

We acknowledge support from CONICET, FONCYT BID 802/OC-AR PICT: 00223 and
SeCyT-UNC.


\end{document}